 \documentclass[nolinenumbers]{aastex63}

\def\etal{et al.~}

\def\gapprox{\lower.4ex\hbox{$\;\buildrel >\over{\scriptstyle\sim}\;$}}
\def\lapprox{\lower.4ex\hbox{$\;\buildrel <\over{\scriptstyle\sim}\;$}}
\def\ref#1{\par\noindent\hangindent1cm {#1}}

\begin{document}

\title{	   The Poissonian Origin of Power Laws in 
	   Solar Flare Waiting Time Distributions }
 
\correspondingauthor{Markus J. Aschwanden}

\author{Markus J. Aschwanden}
\email{aschwanden@lmsal.com}
\affiliation{Solar and Stellar Astrophysics Laboratory (LMSAL),
 Palo Alto, CA 94304, USA}

\author{Jay R. Johnson}
\author{Yosia I. Nurhan}
\affiliation{Andrews University, Berrien Springs, MI 49104, USA}

\begin{abstract}
In this study we aim for a deeper understanding of the power law 
slope, $\alpha$, of waiting time distributions. Statistically 
independent events with linear behavior can be characterized 
by binomial, Gaussian, exponential, or Poissonian size 
distribution functions. In contrast, physical processes with 
nonlinear behavior exhibit spatio-temporal coherence (or memory) 
and ``fat tails'' in their size distributions that fit power 
law-like functions, as a consequence of the time variability of 
the mean event rate, as demonstrated by means of Bayesian
block decomposition in the work of Wheatland et al.~(1998).

In this study we conduct numerical simulations of waiting
time distributions $N(\tau)$ in a large parameter space 
for various (polynomial, sinusoidal, Gaussian) event rate
functions $\lambda(t)$, parameterized with an exponent $p$
that expresses the degree of the polynomial function 
$\lambda(t) \propto t^p$. We derive an analytical
exact solution of the waiting time distribution function
in terms of the incomplete gamma function, which is
similar to a Pareto type-II function and has a power 
law slope of $\alpha = 2 + 1/p$, in the asymptotic limit 
of large waiting times. 
Numerically simulated random distributions reproduce this
theoretical prediction accurately. Numerical simulations 
in the nonlinear regime ($p \ge 2$) predict power law slopes 
in the range of $2.0 \le \alpha \le 2.5$. The self-organized 
criticality model yields a prediction of $\alpha=2$. 
Observations of solar flares and coronal mass ejections 
(over at least a half solar cycle) are found in the range of 
$\alpha_{obs} \approx 2.1-2.4$. Deviations from strict power 
law functions are expected due 
to the variability of the flare event rate $\lambda(t)$, and
deviations from theoretically predicted slope values $\alpha$ 
occur due to the Poissonian weighting bias of power law fits.
\end{abstract} 


\section{	Introduction			}
 
Waiting time analysis, a branch of statistical methods to
discriminate between random processes (with linear behavior)
and processes with intermittence, clustering, and memory 
(with nonlinear behavior)
has become a wide-spread industry to study astrophysical data.
The hallmark of uncorrelated random processes is their
characterization by binomial, Gaussian, Poissonian, or 
exponential distribution functions, while nonlinear
avalanche-like processes reveal power law-like distribution
functions. Applications of such diagnostic tests have been explored
in solar flares (Wheatland et al.~1998;  
Boffetta et al.~1999; Wheatland 2000a;
Leddon 2001; Lepreti et al. 2001; Norman et al.~2001; 
Wheatland and Litvinenko 2002; Grigolini et al.~2002;
Aschwanden and McTiernan 2010; Gorobets and Messerotti 2012;
Hudson 2020; Morales and Santos 2020),
in coronal mass ejections (Wheatland 2003; Yeh et al.~2005;
Wang et al.~2013, 2017);
in solar energetic particle events (Li et al.~2014);
in solar wind discontinuities and its intermittent turbulence
({\bf Carbone et al.~2002;} Greco et al.~2009; Wanliss and Weygand 2007),
in heliospheric type III radio burst storms
(Eastwood et al.~2010),
in solar wind switchback events (Bourouaine et al.~2020;
Dudok de Wit et al.~2020; Aschwanden and Dudok de Wit 2021),
in the cyclic behavior of the solar dynamo over millennia
(Usoskin et al.~2017), 
in active and inactive M-dwarf stars (Hawley et al.~2014;
Li et al.~2018),
in the avalanche dynamics of radio pulsar glitches 
(Melatos et al.~2008; 2018), 
in magnetar bursts (Cheng et al.~2020), stellar
gamma-ray bursts (Guidorzi et al.~2015; Yi et al.~2016),
black hole systems (Wang et al.~2015, 2017),
and in plasma physics experiments
(Sanchez et al.~2002).
For a review of waiting time distributions in astrophysics
see textbook Chapter 5 in Aschwanden (2011).

The power law behavior of waiting time (or laminar time intervals) 
has been interpreted in terms of magnetohydrodynamic (MHD)
turbulence (Boffetta et al.~1999; {\bf Carbone et al.~2002;}
Bourouaine et al.~2020), {\bf complementary} to self-organized 
criticality models, {\bf since both types of models can produce
bursty avalanches, intermittency, power law-like size distributions,
and 1/f-spectra (Carbone et al.~1999)}.
Alternative approaches of analyzing flare waiting times
includes Bayesian decomposition 
(Wheatland and Litvinenko 2002; Wheatland 2004; Wheatland and Craig 2006),
information theory and Shannon entropy (Snelling et al.~2020).
Interestingly, the waiting time distribution of pulses
from active black hole systems exhibit the same power
law slopes as measured in solar flares, $\alpha \approx 2.0$,
which indicates some universality (Wang et al.~2015;
Yi et al.~2016).

Besides the diagnostic capability of waiting
time distributions, which is exploited in many of these studies,
there is almost no study that attempts to explain the
numerical value of the power law slope $\alpha$ in a waiting time
distribution function, which represents a statistical 
invariant. We cannot claim
that we understand waiting time distributions as long as
we do not have a physical model that predicts the observed 
power law slopes. Furthermore, it has not yet been
demonstrated whether power laws exist at all 
(Stumpf and Porter 2012)
in waiting time distributions, or what detailed mathematical
function is to be expected. Recent studies identify
Pareto type-II functions to fit waiting time distributions 
(Aschwanden and McTiernan 2010; Aschwanden and Freeland 2012;
Aschwanden and Dudok de Wit 2021), which exhibit a power law-like
part in the asymptotic limit only. 

In this Paper we calculate analytical exact and approximative  
solutions of waiting time distributions (Section 2 and
Appendices A, B, and C). We present new analytical solutions 
in terms of the gamma function, the incomplete gamma function, 
the Pareto type-II function, while another exact analytical 
solution was recently found in terms of Bessel functions 
(Nurhan et al.~2021).
We test the analytical solutions with numerical simulations 
of random waiting times (Section 3).
We explore the variability of the event rate function 
$\lambda(t)$, which appears to be a key parameter
in the calculation of the power law slope $\alpha$ of 
waiting times, by performing a parametric study of the
power law slope $\alpha$ and nonlinearity parameter $p$ of
the event rate evolution (Section 3).
In the discussion section we juxtapose observed,
theoretical, and numerical waiting time distributions,
examine the existence of power laws and their expected
deviations, as well as their role in self-organized 
criticality models (Section 4). We complete the paper
with a summary of conclusions (Section 5).

\section{	Theory				}

\subsection{	Stationary Waiting Time Distributions }

Statistical probability distributions $N(\tau) d\tau$ of 
event waiting times, $\tau$, have distinctly different 
mathematicl functions, depending
on whether their generation is produced by a linear or
by a nonlinear process. If statistical events are produced 
by a linear process, the resulting size distribution $N(x)\ dx$  
or waiting time distribution $N(\tau)\ d\tau$ follows
Poissonian statistics, which can be fitted by an exponential 
function,
\begin{equation}
	N(\tau)\ d\tau = \lambda_0 \ \exp{ \left[ - \lambda_0
	\tau \right]} \ d\tau \ ,
\end{equation}
where $N(\tau) d\tau$ is the number of events per bin $d\tau$,
$\lambda_0=1/\tau_0$ is the mean event rate
or mean reciprocal waiting time, i.e., $\tau_0 = T/N_0$,
with $T$ being the total duration of an observed data set, and
$N_0$ is the total number of events detected during this
time interval. The total number of events is $N_0$,
\begin{equation}
	\int_0^\infty N(\tau) d\tau = N_0 \ .  
\end{equation}

In contrast, if a statistical sample is produced by
a nonlinear process, the ``fat-tail'' of their distribution
function becomes important and the resulting size distribution 
or waiting time distribution $N(\tau)$ fits a power law distribution 
function,
\begin{equation}
	N(\tau)\ d\tau \propto \ \tau^{-\alpha} \ d\tau \ ,
\end{equation}
with power law slope $\alpha$. This formalism is
valid for a large variety of nonlinear events, including
solar and stellar flares, terrestrial catastrophes,
also known as avalanche events in the parlance of self-organized 
criticality systems. 

How do we distinguish between stationary and non-stationary
statistics? Stationary Poisson processes have a constant
event rate function $\lambda_0$ 
within the statistical uncertainties
in any time interval of a considered data set. Non-stationary
data samples have a variable mean event rate $\lambda(t)$
as a function of time $t$ (also called flare rate function). 
In the case of solar flares, for instance, 
the flare event rate varies by factors of
$\approx 10^2-10^3$ between the solar cycle ($T \approx 11$ years)
minimum and maximum. In the following we will
use the terms {\sl event rate function} and {\sl flare rate
function} $\lambda(t)$ interchangeably.

\subsection{	Non-Stationary Waiting Time Distributions }

We define non-stationary waiting time distribution functions
with a time-dependent flare rate function $\lambda(t)$, which is
a generalization from the constant $\lambda_0$ (Eq.~1) to a 
time-dependent mean flare rate $\lambda(t)$,
\begin{equation}
	N(\tau )\ d\tau = \lambda(t) \ \exp{[-\lambda(t)\ \tau]} \ d\tau \ .
\end{equation}
Subdividing a waiting time distribution into a piece-wise constant
Poisson process with rates $\lambda_i$ and time intervals $\tau_i$,
it can be approximated by (Wheatland 2000b; 2001),
\begin{equation}
	N(\tau_i) \approx \sum_i \varphi_i \lambda_i 
	\exp{[-\lambda_i \tau_i]} \ ,
\end{equation}
where
\begin{equation}
	\varphi_i = {\lambda_i \tau_i \over \sum_i \lambda_i \tau_i } \ ,
\end{equation}
is the fraction of events associated with a given rate $\lambda_i$.
Inserting the fractions $\varphi_i$ (Eq.~6) into the waiting time probability
distribution function (Eq.~5) we obtain,
\begin{equation}
	N(\tau_i) \approx {\sum_i \lambda_i^2 \ \tau_i \ 
		\exp{[-\lambda_i \tau_i]}
	\over \sum_i \ \lambda_i \ \tau_i} \ ,
\end{equation}
which in the asymptotic limit of arbitrary small Bayesian time intervals
approaches the integral $N(\tau)$,
\begin{equation}
	N(\tau) = {\int_0^{T} \lambda(t)^2 \ \tau \ 
		\exp{[-\lambda(t) \tau ] \ dt}
	\over \int_0^T \ \lambda(t) \ \tau \ dt} 
	        = {\int_0^{T} \lambda(t)^2 \ 
		\exp{[-\lambda(t) \tau ] \ dt}
	\over \int_0^T \ \lambda(t) \ dt} \ .
\end{equation}
The right-hand side equation simplifies somewhat with the
cancellation of the variable $\tau$ in both the numerator and
denominator.

{\bf Note that our treatment of non-stationary Poisson processes
follows the previous work of Wheatland (2000b; 2001), which is
related to the concepts of ``mixed Poisson'', 
``compound Poisson'', ``contagion Poisson'', ``Poisson 
autoregression'', and ``Poisson point'' processes
(Kingman 1993; Grandell 1997; Streit 2010).} 

\subsection{	Flare Rate Functions		}

In non-stationary waiting time distributions, the temporal variability
of the mean event rate $\lambda(t)$ needs to be defined, in order
to calculate the waiting time distribution $N(\tau)$ (Eq.~8).
In order to minimize the number of free parameters, following
the principle of Occam's razor, we define here non-stationary event 
rate functions $\lambda(t)$ that have three parameters only 
$(T, \lambda_0, p)$. These flare rate functions have:
(i) a total duration $T$ and
a mean event rate $\lambda_0$, (ii) are 
symmetric in the time interval before and after the peak time $t=T/2$, 
and (iii) vary the shape of the (flare) event rate function (as a 
function of time) by a single exponent $p$ (also called degree or
order of polynomial), to which we refer as nonlinearity parameter
also. We define three different flare rate function models,
including the polynomial, sinusoidal, and Gaussian function,
\begin{equation}
	\lambda(t) = \lambda_0 \left\{
	\begin{array}{ll}
	\left({T/2 - |t - T/2| 
	\over T/2} \right)^p 		
	  &  {\rm Polynomial\ model} \\  
	\left(\sin{\left[ \pi {t \over T} \right]} \right)^{p} 
	  &  {\rm Sinusoidal\ model} \\  
	\left(\exp{\left[ - {(T/2-t)^2 \over 2 \sigma^2} \right]}\right)^p
	  &  {\rm Gaussian\ model} 
	\end{array} \right. \ .  
\end{equation}
These time-dependent functions $\lambda(t)$ are suitable
to represent the time variation of a cycle, such as the solar cycle
of $T \approx 11$ years. 
For the parameter $p=1$, the polynomial model mimics
a simple triangular function with linear rise time and decay time, 
for $p=2$ a quadratic function, for
$p=1/2$ a square root function, and for large exponents $p \gg 1$
it asymptotically approaches the Dirac delta function. 
The parameter space of $t$ and $\lambda(t)$ is shown in Fig.~1
for the polynomial flare rate function, which illustrates
that functions $\lambda(t)$ with degree $p=0.1, ..., 10$
cover most of the $\lambda(t)$ parameter space.

Similar time profiles are formulated for the sinusoidal and
Gaussian function (Eq.~9). The Gaussian width $\sigma$ is defined
from the full width at half maximum (FWHM) being equal to the half
duration, i.e., FWHM=T/2, and $\sigma=(T/4)/\sqrt{2 \ln 2}$.  
The selection of three different functional forms (polynomial,
sinusoidal, and Gaussian) allows us to explore systematic
errors in the power law slope of waiting time distributions,
which are generally larger than the formal errors obtained from
least-square fitting. Another motivation for the choice of our
selection of flare rate functions $\lambda(t)$ is the availability
of analytical exact solutions, for the polynomial model 
(see Section 2.4 and Appendix A), for the sinusoidal flare rate 
model (Nurhan et al.~2021), for the Gaussian flare rate model 
(Appendix C), as well as for the Pareto type-II distribution
function (Appendix B).

\subsection{       Exact Solution of Waiting Time Distribution	}

We derive here an analytical exact solution of the waiting time 
distribution $N(\tau) d\tau$ (Eq.~8),
\begin{equation}
        N(\tau) = {2 \int_0^{T/2} \lambda(t)^2 \
                \exp{[-\lambda(t) \tau ] \ dt}
        \over 2 \int_0^{T/2} \ \lambda(t) \ dt} \ ,
\end{equation}
for the general case of a flare rate function $\lambda(t)$,
chosen symmetrically about the peak time $t=T/2$ in
the time range $0 \le t \le T$, in terms of a polynomial 
with degree $p$ (Eq.~9),
\begin{equation}
        \lambda(t) = \lambda_0 \left({ t \over T/2 }\right)^p \ .
\end{equation}
Inserting Eq.~(11) into Eq.~(10) leads to, 
\begin{equation}
        N(\tau, p) = {\int_0^{T/2} \lambda_0^2\ [t/(T/2)]^{2p} \
                \exp{[ -\lambda_0\ [t/(T/2)]^p\ \tau ] \ dt}
        \over \int_0^{T/2}\ \lambda_0\ [t/(T/2)]^p \ dt} \ .
\end{equation}
Substituting the variable $x=t/(T/2)$, with $dx/dt=1/(T/2)$
and $dt=dx (T/2)$, renders Eq.~(12) as,  
\begin{equation}
        N(\tau, p) = {\int_0^{1} \lambda_0^2\ x^{2p} \
                \exp{[-\lambda_0\ x^p \tau ] \ dx}
        \over \int_0^1\ \lambda_0\ x^p \ dx} \ ,
\end{equation}
where the integral limits change from $t=[0,T/2]$ to $x=[0,1]$. 
The integral in the denominator of Eq.~(13) is simply,
\begin{equation}
	\int_0^1 \lambda_0 \ x^p\ dx = {\lambda_0 \over (p+1)} \ ,
\end{equation}
while the integral $I(\tau, p)$ in the numerator of Eq.~(10) is, 
\begin{equation}
        I(\tau, p) = {\int_0^{1} \lambda_0^2\ x^{2p} \
                \exp{[-\lambda_0\ x^p \tau ] \ dx} } \ .
\end{equation}
We change the variable $y=x^p$, or $x=y^{1/p}$, with the
differential $dx=(1/p) y^{(1-p)/p}\ dy$, which yields,
\begin{equation}
        I(\tau, p) = \int_0^{1} {\lambda_0^2 \over p}
	y^{(1+p)/p} \ \exp{[-\lambda_0\ \tau\ y]} \ dy \ .
\end{equation}
In the limit of $\tau \mapsto \infty$, the exponential
factor drops very fast, so that the upper integral limit 
is of order $O(e^{-\beta})$, where $\beta$ is defined by
\begin{equation}
	\beta = \lambda_0 \tau \ ,
\end{equation}
yielding, 
\begin{equation}
        I(\tau, p) = \int_0^{1} {\lambda_0^2 \over p}
	y^{(1+p)/p} \ \exp{[-\beta\ y]} \ dy \ .
\end{equation}
Substituting $\beta$ with $q=\beta\ y$ and $dq=\beta\ dy$,
\begin{equation}
        I(\tau, p) \approx {\lambda_0^2 \over p\ \beta^{2+1/p}}
	\int_0^\infty\ q^{(1+p)/p} \ \exp{[-q]} \ dy \ ,
\end{equation}
leads to the gamma function $\Gamma$ (Bronstein and Semendjajew 1960),
\begin{equation}
        \Gamma \left( \alpha \right)
        = \int_0^\infty
        q^{(\alpha - 1)} \ \exp{[-q]} \ dq \ ,
\end{equation}
and the integral over the range $q=[0, \infty]$ is, 
in the asymptotic expansion of the exact solution,
\begin{equation}
        I(\tau, \alpha ) \approx {\lambda_0^2 \over p\ \beta^{\alpha}}
	\ \Gamma\left({\alpha}\right) \ ,
\end{equation}
with the power law slope $\alpha$, 
\begin{equation}
	\alpha = 2 + {1 \over p} \ .
\end{equation}
Combining this expression with the denominator integral 
$\lambda_0/(p+1)$ (Eq.~14), we obtain an approximation of the
waiting time distribution function, 
\begin{equation}
        N(\tau, \alpha) \approx \lambda_0 
	\left( {\alpha-1} \right) \
	\Gamma{\left( \alpha \right)} \  
	{(\lambda_0\ \tau)}^{-\alpha} \ . 
\end{equation}
The exact solution of this integral is the incomplete gamma function 
(Bronstein and Semendjajew 1960),
\begin{equation}
	\gamma \left( \alpha, \beta \right) 
	= \int_0^\beta 
	q^{(\alpha - 1)} \ \exp{[-q]} \ dq \ ,
\end{equation}
where the integral boundaries are $q=[0, \beta]$, in contrast
to $q=[0, \infty]$ in the gamma function.
The exact solution of the waiting time distribution
as a function of the waiting time $\tau$ and nonlinearity
parameter $p$ of the flare rate function $\lambda(t) \propto t$
reads then in the final form as,
\begin{equation}
        N(\tau, \alpha) = \lambda_0 \left({\alpha - 1}\right) 
	\gamma\left( \alpha, \lambda_0 \tau \right) \  
	(\lambda_0 \tau)^{-\alpha} \ .
\end{equation}
The power law slope $\alpha=(2+1/p)$ can approximately 
be expressed in terms of the gamma function $\Gamma$,
but the exact solution has an additional dependence of the waiting
time $\tau$ in the incomplete gamma function $\gamma$ (Eq.~25).

Alternative solutions of the waiting time distribution $N(\tau)$
can be obtained also for Pareto Type-II distribution functions,
which is presented in Appendix B. All solutions, the Gamma function 
$\Gamma$ solution (Eq.~21), the incomplete Gamma function
$\gamma$ solution (Eq.~25), and the Pareto type-II distribution
exhibit the same characteristics: The steepest power law slope is
in the asymptotic limit of the largest waiting times, and a
gradual rollover and flattening to a constant value occurs at the
shortest waiting times, as illustrated in Fig.~2.

\subsection{    Self-Organized Criticality Model       }

There are very few studies that provide a prediction or
physical model for the power law slope $\alpha$ of waiting 
time distributions. One of them is
the {\sl fractal-diffusive avalanche model of
a slowly-driven self-organized criticality (FD-SOC) system}
(Aschwanden 2012, 2014; Aschwanden and Freeland 2012;
Aschwanden et al.~2016), which has 
has been further developed from the original version
of SOC concepts (Bak et al.~1987; 1988). It is based
on a scale-free (power law) size distribution function
$N(L)$ of avalanche (or flare) length scales $L$,
\begin{equation}
        N(L)\ dL \propto L^{-d}\ dL \ ,
\end{equation}
with $d$ the Euclidean spatial dimension (which can have
values of d=1, 2, or 3), defining a reciprocal relationship
between the spatial size $L$ 
and the occurrence frequency $N(L)$. 

In the FD-SOC model, the transport process of an avalanche 
is described by classical diffusion,
which obeys the scaling law,
\begin{equation}
        L \propto T^{\beta/2} \ ,
\end{equation}
with $\beta=1$ for classical diffusion. Substituting
the length scale $L \propto T^{\beta/2}$ with
the duration $T$ of an avalanche event, using Eq.~(26-27)
and the derivative $dL/dT = T^{\beta/2-1}$, predicts
a power law distribution function for the size distribution
of time durations $T$,
\begin{equation}
        N(T)\ dT = N(T[L]) \left({dL \over dT}\right) \ dT
                 \ \propto\ T^{-[ 1+ (d-1) \beta/2] } dT
                 \ \propto\ T^{-\alpha_\tau}\ dT
                 \ \approx\ T^{-2} dT \ .
\end{equation} 
For standard parameters $d=3$ and $\beta=1$, it defines the 
waiting time power law slope $\alpha_\tau$,
\begin{equation}
        \alpha_\tau = 1+ (d-1) \beta/2 = 2  \ .
\end{equation}
We can now estimate the size distribution of
waiting times by assuming that the avalanche durations
represent upper limits to the waiting times $\tau$ during
flaring time intervals, while the waiting times become
much larger during quiescent time periods. Such a bimodal
size distribution with a power law slope of $\alpha_\tau \lapprox 2$
at short waiting times ($\tau \le \tau_e$),
and an exponential-like cutoff function
at long waiting times $(\tau \ge \tau_e)$ is approximately,
\begin{equation}
        N(\tau )\ d\tau = \left\{ \begin{array}{ll}
        \tau^{-2}                    & {\rm for}\ \tau \ll \tau_e \\
        \tau^{-2} \exp(-\tau/\tau_e) & {\rm for}\ \tau \gapprox \tau_e \\
                  \end{array} \right. \ .
\end{equation}
Thus, this FD-SOC model predicts a power law with a slope of
$\alpha \approx 2.0$ in the inertial range, and a steepening
cutoff function at longer waiting times. 

\section{	Numerical Simulations and Analysis		}

\subsection{    Power Law Fitting Method 			}

{\bf In this study we employ numerical simulations of 
waiting time distributions, using a random generator, 
a sorting algorithm, and a standard least-square minimization 
algorithm from the {\sl Interactive Data Language (IDL)} software.}
For a stationary waiting time distribution we generate a random
sample of $N_0$ values $t_i$, $i=1,...,N_0$, that are equally
distributed in a unity time interval $[0,1]$.
Then we run a sorting algorithm that sorts the values $t_i$
in a monotonically increasing rank order,
\begin{equation}
        0 < t_1 < t_2 < ... < t_N
\end{equation}
from which a series of $(N_0-1)$ positive time intervals $\tau$ can be
obtained,
\begin{equation}
        \tau_i = (t_{i+1} - t_i) \ , \quad i=1, 2, ..., N_0-1 \ .
\end{equation}
We sample the waiting times $\tau_i$ in a log-log histogram
(Eq.~3) and fit the power law slope $\alpha$ in a suitable inertial 
(scale-free) range, using a standard least-square optimization
algorithm,
\begin{equation}
        \chi =
        \sqrt{ {1 \over (n_{bin} - n_{par})}
        \sum_{j=1}^{n_{bin}}
        {[N^{theo}(x_j)-N^{obs}(x_j)]^2
        \over \sigma_j^2 }
        } \ ,
\end{equation}
where $x_j, j=1,...,n_j$ are the counts per bin width,
$N^{theo}(x_j)$ is the number counts of the theoretical model,
$N^{obs}(x_j)$ is the observed number counts,
$n_{bin}$ is the number of the histogram bins,
and $n_{par}$ is the number of free parameters
of the fitted model functions.
The estimated uncertainty of counts per bin,
$\sigma_j$, is according to Poisson statistics, 
\begin{equation}
        \sigma_j = {\sqrt{N_j \Delta x_j} \over \Delta x_j} \ ,
\end{equation}
where $\Delta x_j$ is the (logarithmic) bin width.
The goodness-of-fit $\chi$ quantifies which theoretical
model distribution is consistent with the (observed) data.
The uncertainty $\sigma_\alpha$ of the best-fit power law
slope $\alpha$ is estimated from
\begin{equation}
        \sigma_\alpha = {\alpha \over \sqrt{N_0}} \ ,
\end{equation}
with $N_0$ the total number of events in the entire size
distribution (or in the fitted range).

The log-log histogram of the waiting times $\tau$ is sampled
with a resolution of $n_{bin}=6$ bins per decade, typically
for a sample of $N_0 \approx 10^4$ events. The power law
fit is performed with a least-square fit in the range of
$[x_0, x_2]$, where the lower bound $x_0$ is defined by the
maximum of the log-log waiting time histogram, and a margin
of $d_{bin}=1$ bin above $x_0$ is ignored in the power law fit
in order to minimize the influence of the roll-over in the
least-square fit of the power law slope. The upper bound $x_2$
of the inertial range is given by the largest waiting time.
The data analysis performed here employs identical methods as
in other studies (Aschwanden and McTiernan 2010;
Aschwanden 2015, 2019a, 2019b; Aschwanden and Dudok de Wit 2021).

{\bf It is widely known that fitting on a log-log scale is biased
and inaccurate, while using a maximum likelihood estimation 
is more robust (Goldstein et al.~2004; Newman 2005; Bauke 2007).
We take this into account by using the Poissonian error 
estimate (Eq.~35). However, while this improves the formal error, 
there is a much larger systematic error due to deviations 
from ideal power laws (Stumpf and Porter 2012). One of the major
goals of this paper is to pin down such systematic errors by
calculating the exact analytical solutions of waiting time
distributions, which are generally not ideal power laws,
but rather convolutions with the (incomplete) Gamma function.}

\subsection{	Analytical Solution Versus Numerical Simulation	}

In principle, there are three different methods of calculating
a waiting time distribution: (i) by an analytical solution, (ii) 
by numerical integration of the analytical integral equation
(Eq.~8), and
(iii) by sampling of numerically simulated events, which essentially
is a Monte Carlo simulation method. The problem is posed by
the integral equation of the probability distribution function 
$N(\tau)$ as a function of the waiting time $\tau$ (Eq.~8) and 
a chosen flare rate function $\lambda(t)$ (for instance the
flare rate functions given in Eq.~9). 

We juxtapose the numerically simulated waiting time distributions
(histogram in Fig.~2), the power law fit in the inertial range 
of this histogram (dashed line in Fig.~2), the exact
analytical solution according to Eq.~(A6) given in Appendix A
(thick solid curve in Fig.~2), 
and the Pareto type-II approximation Eq.~(B3) 
(thin solid curve in Fig.~2).
We see that the power law slope of the exact analytical
solution, $\alpha_{exact}=3.00$, is significantly steeper than
the slope of the power law fit to the simulated data, $\alpha
=2.67\pm0.03$, by a factor of $\alpha_{exact}/\alpha_{fit}=1.12$
(or 12\%). This occurs because of the Poissonian weighting bias,
which produces the smallest uncertainties $(\sigma \propto \sqrt{N})$
in the bins at the lower bound of the inertial range (at $x_0$), 
and thus constrains the power law fit there most. 
We found that relatively small
samples with $N_0 \approx 10^3$ events provide a better overall fit
(according to the goodness-of-fit criterion Eq.~33),
rather than large samples of order $N_0 \gapprox 10^5$, for the same
reason (of the weighting by Poisson statistics). The formal error from
$\chi$-square fitting is typically $\approx 1\%$ for a sample size
of $N_0 \approx 10^4$, according to Poisson statistics (i.e.,
$\sigma \propto \sqrt{N_0}$), which is generally smaller than
the systematic errors (in the order of $\gapprox 10\%$). 

\subsection{	Multi-Poissonian Waiting Time Distributions	}

We show the results of some numerical simulations in Figs.~(3) and (4). 
In Fig.~(3) we show six simulation runs with varying temporal resolution.
In the first case (Fig.~3a) the time resolution is equal to the total
duration of the observations, $\Delta t=T$, which corresponds to a 
constant flare rate $\lambda(t) =\lambda_0$, yielding a perfect 
exponential waiting time distribution (Fig.~3g), as defined in Eq.~(1).
When we increase the resolution to  $\Delta t=T/2$ (Fig.~3b), we see
that the waiting time distribution $N(\tau)$ exhibits a double hump
(Fig.~3h), which is the superposition of two flare rates, one at 
the solar cycle minimum and one at the solar cycle maximum. This
double hump structure persists on a weaker level when we increase
the time resolution to $\Delta t = T/4$ and T/8 (Figs.~3i and 3j), 
but is morphing
towards a perfect power law distribution for $\Delta t = T/16$ and T/32
(Figs.~3k and 2l). This series of simulations demonstrates most
clearly the Poissonian origin of power laws. Essentially, broadening
the flare rate $\lambda(t)$ transforms the waiting time distribution 
$N(\tau)$ from an exponential (Eq.~1) to a power law distribution 
function (Eq.~3). The theoretical slope of a continuous sinusoidal
model with $\lambda(t) \propto \sin{(t)}^2$ with exponent
$p=2$ is $\alpha=2+1/p=2.5$ (Eq.~22), which agrees well with the simulations
with $\alpha \approx 2.5\pm0.1$ (Fig.~3).

\subsection{	Variation of Flare Rate Functions	}

In Fig.~4 we show three examples of simulated shapes of
flare rate functions $\lambda(t)$, selecting one example 
(with parameter $p=1$) from each of the three
(polynomial, sinusoidal, and Gaussian) model groups.
The first case is generated by a linear rise time and decay time in the 
flare rate function $\lambda(t) \propto t$, which is simply a polynomial
function with exponent $p=1$. The time evolution, analytically
defined by Eq.~(9), is shown in Fig.~(4a), 
and the resulting waiting time distribution $N(\tau)$ is 
approximated by a power law slope with value $\alpha=2.67\pm0.03$ 
(Fig.~4d), in an inertial range of $\tau \approx 10^{-3}, ..., 10^{-1}$,
from a sample of $N_0=1.0 \times 10^4$ events. The same fit is shown
in Fig.~2, juxtaposed to the analytical solution and the Pareto type-II
approximation. The total duration of the simulation 
is $T=11$ years (to mimic the solar cycle), and the
number of temporal Bayesian blocks in the flare rate function
is chosen in $N_t=120$ time bins here, corresponding to a bin width of
$dt=T/N_t \approx 0.1$ (years). Similarly, the flare rate time
profiles $\lambda(t)$ is shown for a sinusoidal (Fig.~4b)
and for a Gaussian model (Fig.~4c). 
The various models of flare rate functions $\lambda(t)$ shown 
in Fig.~4 illustrate that they all produce
power law-like waiting time distributions $N(\tau)$ with slopes
that are almost identical, in the range of $\alpha \approx 2.66-2.70$,
for a nonlinearity parameter of $p=1$.  
Hence, the power law slopes $\alpha$ are not very sensitive
to the detailed shape of the flare rate functions $\lambda(t)$,
as long as a similar range of flare rates is covered. 
The Gaussian event rate function, however, has a finite value
at $t=0$ and $t=T$, which affects the power law slope somewhat,
compared with the polynomial and the sinusoidal case that have 
zero-values, $\lambda(t=0)=0$ and $\lambda(t=T)=0$.

\subsection{	Power Law Slopes of Waiting Time Distributions	    }

Since we are interested in quantifying the waiting time distribution
$N(\tau) \propto \tau^{-\alpha}$ from first principles, we need
a relationship between the power law slope $\alpha$, which is an
observable, and the free parameter $p$ that characterizes the 
flare rate function $\lambda(t) \propto t^p$ (Eq.~9). We apply
linear regression fits and show the values of 
the best-fit power law slopes $\alpha(p)$ as a function of the 
flare rate function exponent $p$ in Fig.~(5), for a range of 
p=0.5 - 6.0 (tabulated in Table 1). Each value of the 36 runs 
corresponds to a separate numerical simulation.
We find that the relationship 
between the two parameters $\alpha$ and $p$ can be determined by  
linear regression fits (Fig.~5),
\begin{equation}
	\alpha(p) \approx \left\{
	\begin{array}{ll}
	2.09 + 0.65 /p & \quad {\rm Polynomial\ model} \\
	2.31 + 0.60 /p & \quad {\rm Sinusoidal\ model} \\
	2.12 + 0.70 /p & \quad {\rm Gaussian\ model} \\
	\end{array} 
	\right. \ , 
\end{equation}
Note that we applied the correction factor 
$\alpha_{corr}/\alpha_{fit}=1.12$ to the observed (or
numerically simulated) power law slopes $\alpha_{fit}$ 
due to the Poissonian weighting bias (see Section 3.2 and Fig.~2).
These best-fit relationships reproduce the chief features of
the theoretical prediction ($\alpha = 2 + 1/p$),
based on the gamma function $\Gamma$ (Eq.~23) in the 
approximative case, or based on the incomplete gamma
function $\gamma$ (Eq.~25) in the exact analytical solution:
(i) An asymptotic value of $a \approx 2.0$ is obtained in the 
nonlinear regime ($p \mapsto \infty$), and (ii) an
asymptotic value of $\alpha \gapprox 2$ is reached in the linear regime 
($p \lapprox 1$). 
The polynomial flare rate relationship is shown in Fig.~(5a), 
while similar values were obtained for the sinusoidal (Fig.~5b) 
and the Gaussian flare rate model (Fig.~5c). 
An exact solution for a sinusoidal flare rate function
including a full period
and its waiting time distribution is given in Nurhan et al.~(2021)
with $\alpha=2.5$.

In practice, the power law slope $\alpha_{pred}$ can
now be predicted from the observed parameter $p_{obs}$ in the flare
rate function $\lambda(t) \propto t^p$, 
\begin{equation}
	\alpha_{pred} = 2 + {1 \over p_{obs}} \ ,
\end{equation}
or vice versa, the nonlinearity parameter $p_{pred}$ can be 
predicted from from fitting the power law slope $\alpha_{obs}$,
\begin{equation}
	p_{pred} = {1 \over \alpha_{obs} - 2 } \ .	
\end{equation}

Solar flare observations reveal a small range of power law
slopes $\alpha \approx 2.1-2.4$ 
(Boffetta et al.~1999; Wheatland 2003; Wheatland and Litvinenko
(2002), Wheatland 2000a; Lepreti et al.~2001; Grigolini 
et al.~2002; Aschwanden and McTiernan 2010), which map
onto a polynomial exponent $p$ in the range of $p \approx 2.5-7.0$
(Fig.~6; Table 2). This range is expected since nonlinear processes
require ($p \gapprox 2$). 

\subsection{		Randomized Flaring 		}

So far we modeled the time variability of the waiting time 
distribution on the largest scale of a full cycle, which has
a typical time duration of $T \approx 11$
years, which we sampled with a time resolution of $dt = 0.1$ year 
(or 36 days).
In reality, however, there are temporal structures on much
shorter time scales, down to $\tau \approx 1.0$ minute during
episodes of high flaring activity, which strongly deviates
from the slowly-varying solar cycle modulation of 11 years
assumed in our previous modeling of the flare rate function
$\lambda(t)$. 

We perform an experiment of simulating the stochasticity of 
solar flare rates by random shuffling of the flare times $t_i$.
The time profile $\lambda(t)$ shows a linear (positive) increase 
and a linear (negative) decrease in our first model (Fig.~7a).  
From this slowly-varying polynomial flare rate function with
24 time bins we shuffle some of the time bins arbitrarily,
which mimics an irregular, stochastic, fast-varying
flare rate function $\lambda(t)$ (Fig.~7b). As expected,
the resulting waiting time distribution $N(\tau)$ reveals 
an identical power law slope, i.e., $\alpha \approx 2.2$
for both the slowly-varying and fast-varying polynomial functions
over the duration of a solar cycle. This outcome is trivial, 
because it simply reflects the commutative property of arithmetic
sums in linear algebra: The sum of the time-varying waiting 
time distributions is invariant to the time order. For instance,
$\sum x_1 + x_2 + x_3 = \sum x_3 + x_1 + x_2$, or any other
permutation of values $x_i$. The commutative property 
simplifies our modeling enormously, since a simple monotonic
increase of flare rates includes all possible permutations also.

A corollary of the communtative property is that the power law
slope $\alpha$ is independent of the time duration of the
observational sample or the time resolution, as long as the
same range of flare rates is covered during the sample.
Consequently, an observational sample may contain 
intermittent flaring of many short-duration flare bursts
and still produce power law slopes of waiting times
that are similar to those samples with a slowly-varying 
solar cycle profile. An illustration of a time profile containing
multiple bursts is depicted in Fig.~7c, which again produces 
the same power law slope ($\alpha=2.23\pm0.02$; Fig.~7f) 
as the entire solar cycle ($\alpha=2.23\pm0.02$; Fig.~7d).

\section{	Discussion 				}

\subsection{	Observational Constraints of Waiting Time 
		Distributions 	}

A comprehensive list of published waiting time distributions
of solar flares has been presented in Table 1 of Aschwanden 
and Dudok de Wit (2021). In Table 2 of this study here
we compile a subset of these observed waiting time distributions 
that cover at least a half solar cycle ($T/2 \gapprox 5$ years),
so that they contain the full flare rate variability
from the solar minimum to the solar maximum (or vice versa).
This selection criterion contains solar flares observed with 
the Hard X-ray Burst Spectrometer onboard the Solar Maximum Mission
(HXRBS/SMM), the Rueven Ramaty High Energy Solar Spectrometric Imager
(RHESSI), 
the Burst and Transient Source Experiment onboard the
Compton Gamma Ray Observatory (BATSE/CGRO), 
the Geostationary Orbiting Earth Satellite (GOES), 
as well as coronal mass ejections observed with 
the Large Angle Solar Coronagraph onboard the
Solar and Heliospheric Observatory (LASCO/SOHO).
Large flare rate variability warrants power law functions
for waiting time distributions, while insignificant
flare rate variability renders exponential (stationary
Poissonian) waiting time distributions.

What power law slopes $\alpha$ of waiting time distributions
have been observed during full (or half) solar cycles ?
As we show in Fig.~6, the combined range of observations
narrows down to a relatively small range of power law slopes 
$\alpha_{obs} \approx 2.1-2.4$. These power law slopes $\alpha$
correspond to a range of $p\approx 2.5-7.0$ for the degree $p$ 
of the polynomial flare rate function (Fig.~6), according to 
the predicted relationship of $\alpha=2+1/p$ (Eq.~22). 
This means that the rise time of the flaring rate is 
not linear ($p \approx 1$), but rather is highly nonlinear 
($p \gapprox 2$). Following this interpretation,
all observed data sets are consistent with nonlinear behavior
(Boffetta et al.~1999; Wheatland 2003,  
Wheatland and Litvinenko 2002; Wheatland 2000a; Lepreti et 
al.~2001; Grigolini et al.~2002). The nonlinear behavior 
is expected for solar flares 
because of its (nonlinear) avalanche-like growth characteristics 
that is typical for self-organized criticality models. 
The implication of this result stretches
much further than to the rise time characteristics of the
solar cycle, because it applies also to the average
evolutionary behavior of many (clustered and intermittent) 
flare rate bursts, as depicted in Fig.~(7c). 

Interestingly, a power law slope of $\alpha = 2.5$ has also
been analytically derived for the case of the sinusoidal flare rate
function, in terms of Bessel functions (Nurhan et al.~2021),
which is identical to our derivation of the lower limit
$\alpha \le 2+1/p = 2.5$ for nonlinear processes ($p \ge 2$).
Theoretically, the power law slope is mostly determined by
the behavior of the flare event rate function $\lambda(t)$
at the times with the slowest event rate, near $t=0$ and $t=T$,
where $\lambda(t) \propto \sin{(\pi t/T)}^2$, and thus 
$\alpha=2+1/p=2.5$ for $p=2$. 

\subsection{	Do Power Laws Exist ?			}

In this study we use numerical tools that can produce
true random distributions of flare events (which should
follow exponential functions with Poisson statistics),
as well as non-stationary Poissonian distributions.
The latter group produces more or less power law-like
functions, but it has never been demonstrated whether
we should expect an exact power law function. Since 
numerous deviations from exactly straight power law 
functions (in a log-N log-S histogram) have been noted
before (Wheatland et al.~1998; Aschwanden and McTiernan 2010),
some criticism has been raised whether power laws
exist at all (Stumpf and Porter 2012). Here we have an
answer that is summarized in Fig.~2. We calculated the
analytical exact solution of a waiting time distribution
for the linear case (with time evolution 
$\lambda(t) \propto t^p = t^1$,
or $p=1$), which is derived in Appendix A, Eq.~(A6). 
The solution contains terms of 
$\tau$, $\tau^2$, $\tau^3$, and $e^{-\tau}$,
and therefore is of the type of an ideal power law function
$\tau^{-\alpha}$ in the asymtotic limit only
($\lambda_0 \tau \mapsto \infty$), where all other terms
are exponentially small.
A Pareto type-II function
$(1 + \tau)^{-\alpha}$ is a good approximation
(Aschwanden and Dudok de Wit 2021; 
Aschwanden and McTiernan 2010; Aschwanden and Freeland 2012)
that is accurate to first-order terms,
but shows a gradual change in the power law slope with
respect to the exact analytical solution. Moreover, in
our numerical simulation of waiting time distributions
we obtain a power law-like function too (Fig.~2),
but the Poissonian weighting of a power law fit is
biased towards the lower end (where the error bars are
smallest and thus have the largest weight there). Thus,
if a power law fit is obtained over the full inertial
range, we predict that the average slope is slightly flatter
than the asymptotic limit at the upper end, which is
evident in the example shown in Fig.~2 (dashed
line). Other examples are shown in Fig.~3, where
deviations from ideal power laws show up clearly,
since the non-stationary Poissonian model (with
large time variability of the mean flare rate) 
predicts that a power law-like convolution consists
of a superposition of multiple (but finite number) 
of exponential distributions.
In summary, deviations from ideal power law functions
are not significant for relatively small samples
(with $N_0 \approx 10^1-10^3$), but become significant
for larger samples (with $N_0 \gapprox 10^4$), as
demonstrated in other works (Aschwanden 2015,
2019a, 2019b, Aschwanden and Dudok de Wit 2021), 
which are theoretically expected,
and thus should be modeled correspondingly.

\subsection{	Self-Organized Criticality Models	}

After the previous disgression to previous observations
and the discussion of whether power laws exist at all,
let us come back to physical models that possibly could
explain the observed power law slopes $\alpha$ of
waiting time distributions.

Naively, to first order, one would expect that the
distribution $N(\tau)$ is reciprocal to the mean waiting
time, i.e., $N(\tau) \propto \tau^{-1}$, simply because
the number of fragments is reciprocal to the length
of a fragment in a one-dimensional ``fragmentation'' process.
In other words, the product of the number $N$ multiplied 
with the duration of a time scale $T$ is invariant, i.e.,
$N_T \times T$ = const. However, this simplest
model that predicts a scaling of $\tau^{-1}$ is not
consistent with observations, which show distributions of 
$\approx \tau^{-2}, ..., \tau^{-3}$.
It is likely that the one-dimensionality in the time
domain is the culprit in this oversimplified model.

However, if we consider the generation of
waiting times in the spatio-temporal domain, a spatial
scale with length $L$ and time scale $T$ are required
in a minimal model, which are coupled by the length
scale distribution $N(L) dL$ and the transport equation
$L(T)$. We can find such a model in the self-organized
criticality concept, as briefly summarized in Section 2.5.
Both relationships can be expressed with power laws,
the scale-free size distribution of avalanches,
$N(L)\ dL \propto L^{-d} dL$ (Eq.~26), and the diffusive 
transport mechanism, $L \propto T^{\beta/2}$ (Eq.~27), with
$d=1,2,3$ the Euclidean dimension, and $\beta=1$ the
classical transport coefficient. Taken these two
relationships together, our SOC model predicts a
power law distribution of flare durations $T$, i.e.,
$N(T) \propto T^{-\alpha}$, where the power law
slope is defined as $\alpha = 1 + (d-1)\beta/2$ (Eq.~28).
Thus, the SOC model predicts a value of $\alpha=2$
for the 3-D space geometry, while the values of
$\alpha=1$ for 1-D and $\alpha=1.5$ for 2-D geometry 
are below the observed values, and thus can be ruled out.
Interestingly, the predicted value of $\alpha=2.0$ for
the 3-D world agrees with the independently calculated
approximation of the Pareto type-II model (Eq.~B2) in the
asymptotic limit of $\tau \mapsto \infty$ (Eq.~B3),
which corroborates the SOC theory and the Pareto 
distribution function to some extent.  

\section{	Conclusions 			}
 
In this study we explore the origin of waiting time
distribution functions, which provides a means to 
diagnose whether the events of a statistical distribution 
are produced by independent random events (obeying 
an exponential function) or by coherent clusters 
(obeying a power law function). This study is a follow-up
on the pioneering work of Wheatland (2003) who concluded
that non-stationary Poissonian processes produce
power law-like distributions due to the (non-stationary)
time variability of the mean flare rate. 
{\bf Although we find good agreement between 
the simulated and theoretically predicted parameters in the
framework of non-stationary Poisson models postulated by
Wheatland et al.~(1998), the presented results do not rule out
alternative interpretations, such as the nonlinear dynamics
of MHD turbulence in active regions (Boffetta et al 1999;
Carbone et al.~2002).}
Our findings and conclusions are briefly summarized in the following:

\begin{enumerate}

\item{The theoretical calculation of a waiting time
distribution function $N(\tau)\ d\tau$ represents a 
convolution of individual exponential distribution 
functions with a time-dependent flare rate function
$\lambda(t)$. The flare rate function can be chosen
arbitrarily, for which we choose polynomial, sinusoidal,
and Gaussian functions (see 36 cases in Table 1), but we 
found that all three functions produce almost identical results, 
because their time variability is invariant to the 
multiplicity and temporal permutation of time structures.} 

\item{New analytical solutions of the waiting time
distribution function $N(\tau) d\tau$ have been found
for polynomial flare rate functions $\lambda(t) \propto t^p$
in terms of the incomplete gamma function, as well as
for the sinusoidal flare rate functions $\lambda(t) \propto
\sin(t)^2$ in terms of Bessel functions 
(Nurham et al.~2021). Another useful approximation
to polynomial flare rate functions is the Pareto type-II
distribution function,
$N(\tau)=N_0 \left[1+c\tau/\alpha\right]^{-\alpha}$,
with $\alpha=2+1/p$ being the (waiting time) power law slope.}

\item{The quadratic polynomial flare rate function,
$\lambda(t) \propto t^p$, implies a high nonlinearity
of degree $p \gapprox 2$ in the time variability of 
the flare rate. The nonlinear time variability
applies theoretically to the long-duration variability 
of the solar cycle, but the real data reveal a high
degree of intermittency, so that the nonlinear evolution
applies to short-duration variability of active
regions and to clustered flare burst episodes too.}

\item{We define a nonlinearity parameter $p$ (or
degree of polynomial order) that characterizes the flare rate 
function $\lambda(t) \propto t^p$ in terms of its evolution, 
which has a theoretically predicted relationship of 
$\alpha \approx 2 + 1/p$ between the nonlinearity parameter $p$
and the waiting time power law slope $\alpha$.
Thus we can predict the power law slope $\alpha$ as a function
of the nonlinearity degree $p$. For instance, for linear
processes ($p=1$) we expect a slope of $\alpha_1 = 3.0$,
and likewise, for a nonlinear process ($p \approx 2$) we expect
$\alpha_2=2.5$. Vice versa, we can predict the degree
of nonlinearity $p$ from an observed waiting time
power law slope $\alpha$, using the inverse relationship,
$p=1/(\alpha-2)$.}

\item{A value of $\alpha=2+1/p$ has been predicted for 
the power law slope of waiting time distributions, 
based on analytical exact solutions in terms of the 
incomplete gamma function. The same value could be
retrieved from the Pareto type-II distribution function.
A fixed value of $\alpha=2.5$ has been obtained 
for the sinusoidal flare rate model (Nurhan et al.~2021),
which is consistent with the nonlinear regime ($p \approx 2$)
and would be expected for a process driven by the
magnetic activity cycle of the solar dynamo.
Another prediction of $\alpha=2$ has been derived from a 
self-organized criticality model, assuming a space-filling
Euclidean dimension of ($D=3$) and classical diffusion as
transport process ($\beta=1$), which occur, for instance, 
in chain reactions of solar flare avalanches. The agreement 
between the independent theoretical models corroborates 
our interpretation in terms of SOC models.}

\item{Do power laws exist ? This question is absolutely
justified, as we demonstrate with three slightly different
approximations in the case of waiting time distributions.
Since a waiting time distribution of non-stationary
Poissonian processes represents a superimposition of
exponential functions, one can easily show deviations
from ideal power laws. However, deviations from ideal 
power law functions are generally not significant for 
relatively small samples (with $N_0 \approx 10^2-10^3$), 
but become significant for larger samples (with 
$N_0 \gapprox 10^4$).}

\end{enumerate}

Future studies may investigate the predictions made for
waiting time distributions by means of observational 
data sets, complementary to the numerical simulations 
demonstated here. It would be particulary important
to test the clustering and sympathetic flaring on
shorter time scales than the solar cycle examined here.
A further benefit of this study is that the new analytic
exact solutions for waiting time distributions,
$N(\tau) d\tau$, could also be extended to occurrence
frequency distributions of (flare) event durations, 
$N(T)\ dT$, which always exhibited larger deviations
from strict power laws than the (flare) event size
distributions, $N(S)\ dS$ (Bak et al.~1987, 1988).

\clearpage

\section*{	APPENDIX A: Analytical Solution of 
	  	Waiting Time Distribution for Case p=1	   }

The waiting time distribution $N(\tau)\ dt$ of a Poissonian
random process can be represented by an exponential distribution
(Eq.~1) for the case of a stationary flare rate function 
$\lambda(t)=\lambda_0$.
For a linearly increasing flare rate function $\lambda(t)=t/(T/2)$
(with $p=1$ in Eq.~9), the probability function is (Wheatland
et al.~1998; Wheatland and Litvinenko (2002); Aschwanden and
McTiernan (2010),
$$
	N(\tau) = 
	{\int_0^T \lambda(t)^2 \exp{\left[-\lambda(t) \tau \right]} dt
	\over \int_0^T \ \lambda(t) \ dt } \ .
	\eqno(A1)
$$
Inserting the flare rate function $\lambda(t)=\lambda_0 
({t / (T/2)})^1$ we obtain the integral,  
$$
	N(\tau) = 
	{\int_0^{T/2} \lambda_0^2 \left({t \over T/2}\right)^2 
	\exp{\left[-\lambda_0\ \tau\ (t/(T/2)) \right]} dt
	\over \int_0^{T/2} \lambda_0 \left({t \over T/2}\right) dt} \ .
	\eqno(A2)
$$
Substituting the variables $x=t/(T/2)$, $dt=(T/2) dx$, and
$a=-\lambda_0 \tau$, we obtain the integral,
$$
	N(\tau) = 
	{\int_0^1 \lambda_0^2 \ x^2  
	\exp^{a x} dx
	\over \int_0^1 x \ dx} \ ,
	\eqno(A3) 
$$
which has the analytical solution (Bronstein and Semendjajew 1960),
$$
	N(\tau ) = \lambda_0 \ \exp^{a x}
	\left({x^2 \over a} - {2x \over a^2} + {2 \over a^3} 
	\right)_{x=0}^{x=1}
	\eqno(A4)
$$
$$
	N(\tau ) = 2 \lambda_0 \exp^a
	\left[ \left({1 \over a} - {2 \over a^2} + {2 \over a^3} \right) 
	- {2 \over a^3} \right] \ ,
	\eqno(A5)
$$
which can be expressed with $a = -\lambda_0 \tau$ to show the explicit
function of the waiting time $\tau$,
$$
	N(\tau ) = 2 \lambda_0 
	\left[ {2 \over \lambda_0^3 \tau^3} -
	\exp^{-\lambda_0 \tau}
	\left({1 \over \lambda_0 \tau} + {2 \over \lambda_0^2 \tau^2}
	+ {2 \over \lambda_0^3 \tau^3} \right) \right] \ .
	\eqno(A6)
$$
\section*{    APPENDIX B: Pareto Type-II Approximation           }

An alternative generalization of a simple straight power law
distribution function $N(\tau) \propto \tau^{-\alpha}$ is
the Pareto Type-II distribution function, 
$N(\tau) \propto (1 + \tau)^{-\alpha}$, which exhibits
a power law-like function at large waiting times and 
becomes constant at small waiting times. Such a Pareto type-II 
function has been used previously to fit waiting time 
distributions (Aschwanden and McTiernan 2010; Aschwanden and 
Freeland 2012; Aschwanden and Dudok de Wit 2021). Here we
present an analytical approximation of the waiting time
distribution function.

We start with the expression of the waiting time distribution
in terms of the gamma function (Eq.~23). In the asymptotic limit
we have
$$
	N(\tau )\sim \lambda_0 \ (\alpha -1)\Gamma (\alpha )
	\ (\lambda_0 \tau )^{-\alpha } \ ,
	\eqno(B1)
$$
where the slope $\alpha$ of the power law is (Eq.~22),
$$	
	\alpha =2+{1 \over p} \ .
	\eqno(B2)
$$
This is also asymptotic to a Pareto type-II distribution of the form,
$$
	N_P (\tau )=\frac{N_0 }{{\left(1+\frac{c\tau }
	{\alpha }\right)}^{\alpha } } \ .
	\eqno(B3)
$$
Equating the two expression (Eq.~B1) and (Eq.~B2) yields,
$$
	N_0 \left(\frac{\alpha^{\alpha } }{c^{\alpha } }
	\right)=\lambda_0 \frac{(\alpha -1)\Gamma (\alpha )}
	{(\lambda_0 )^{\alpha } } \ .
	\eqno(B4)
$$
Also, for a uniform approximation, we require that 
$$
	N_P (0)=N_0 (1-c\tau +...)
	\eqno(B5)
$$
so that the constant $N_0$ is,
$$
	N_0 =\lambda_0 \left(\frac{p+1}{2p+1}\right)=\lambda_0 
	\left(\frac{\alpha -1}{\alpha }\right) \ ,
	\eqno(B6)
$$
which then yields the equivalence,
$$
	\lambda_0 \left(\frac{\alpha -1}{\alpha }\right)
	\left(\frac{\alpha^{\alpha } }{c^{\alpha } }\right)=
	\lambda_0 \frac{(\alpha -1)\Gamma (\alpha )}
	{(\lambda_0 )^{\alpha } } \ ,
	\eqno(B7)
$$
and the constant $c$,
$$	
	c=\frac{\alpha \lambda_0 }{{\left(\alpha \Gamma 
	(\alpha )\right)}^{1/\alpha } }
	\eqno(B8)
$$
This theotetical solution has been checked numerically
and compared with the exact solution based on the
incomplete gamma function solution (Eq.~25), 
the asymptotic approximation, the Pareto distribution 
obtained from the power series, and the Pareto distribution 
obtained from the Pad\'{e} approximation.  
The power series solution is inaccurate when $\lambda_0 \tau >1$. 
The Pad\'{e} approximation is reasonably accurate for all 
$\lambda_0 \tau$ and matches the power law slope accurately.

\section*{	APPENDIX C : Solution for Gaussian Event Rate Function	}

After we calculated exact solutions for the polynomial flare rate
function $\lambda(t)$ (Section 2.4), while exact solutions for the
sinusoidal flare rate are given in Nurhan et al.~(2021), we
derive here an approximate solution for the Gaussian flare rate also.
The definition of the waiting time integral is (Eq.~10),
$$
	N(\tau )=\frac{\int_0^{T/2} \lambda^2 (t)
	\exp (-\lambda (t)\tau )dt}{\int_0^{T/2} \lambda (t)dt} \ ,
	\eqno(C1) 
$$
where the Gaussian rate function is (Eq.~9),
$$
	\lambda (t)=\lambda_0 \exp (-(t-T/2)^2 /2\sigma^2 ) \ . 
	\eqno(C2)
$$
We change the variables 
$$
	y=(T/2-t)/\sqrt{2}\sigma \ ,
	\eqno(C3) 
$$
$$
	dy=-dt/\sqrt{2}\sigma \ ,
	\eqno(C4)
$$
$$
	\lambda =\lambda_0 e^{-y^2 } \ ,
	\eqno(C5)
$$
$$
	\beta =\frac{T}{2^{3/2} \sigma } \ ,
	\eqno(C6)
$$
and obtain the waiting time integral in terms of the error function
erf($\beta$),
$$
	N(\tau )=\lambda_0 \frac{\frac{2}{\sqrt{\pi }}\int_0^{\beta } 
	e^{-2y^2 } \exp (-\lambda_0 e^{-y^2 } \tau )dy}{\frac{2}
	{\sqrt{\pi }}\int_0^{\beta } e^{-y^2 } dy}=\lambda_0 
	\frac{I(\tau ,\beta )}{{\mathrm{e}\mathrm{r}\mathrm{f}}(\beta )} \ ,
	\eqno(C7)
$$
where
$$
	I(\tau ,\beta )=\frac{2}{\sqrt{\pi }}\int_0^{\beta } e^{-y^2 } 
	\exp (-\lambda_0 e^{-y^2 } \tau )dy \ ,
	\eqno(C8)
$$
$$
	{\mathrm{e}\mathrm{r}\mathrm{f}}(\beta )=\frac{2}{\sqrt{\pi }}
	\int_0^{\beta } \exp (-x^2 )dx \ ,
	\eqno(C9)
$$
Now we have, 
$$
	I(\tau ,\beta )=\frac{1}{\lambda_0^2 }\frac{d^2 }{d\tau^2 }
	\frac{2}{\sqrt{\pi }}\int_0^{\beta } 
	\exp (-\lambda_0 e^{-y^2 } \tau )dy \ ,
	\eqno(C10)
$$
We expand the exponential around $y=\beta$, as this will be 
the smallest value in the exponential and will dominate asymptotically.
$$
	e^{-y^2 } =e^{-\beta^2 } (1-2\beta (y-\beta )+...) \ ,
	\eqno(C11)
$$
Consequently we obtain,  
$$	\int_0^{\beta } \exp (-\lambda_0 e^{-y^2 } \tau )dy\sim 
	\int_{\beta -\epsilon }^{\beta } \exp (-\lambda_0 \tau 
	e^{-\beta^2 } (1+2\beta (\beta -y))dy=\exp (-\lambda_0 
	\tau e^{-\beta^2 } )\int_0^{\epsilon } \exp (-2\lambda_0 
	\tau \beta e^{-\beta^2 } q)dq \ , 
	\eqno(C12)
$$ 
which introduces only an exponenitally small error, so that we can 
now take $\epsilon \to \infty$,
$$
	\int_0^{\beta } \exp (-\lambda_0 e^{-y^2 } \tau )dy\sim 
	\frac{\exp (-\lambda_0 \tau e^{-\beta^2 } )}{2\lambda_0 
	\tau \beta e^{-\beta^2 } } \ . 
	\eqno(C13)
$$
Taking the leading order terms,
$$
	-\frac{d}{d\tau }\frac{\exp (-\lambda_0 \tau e^{-\beta^2 } )}
	{2\lambda_0 \tau \beta e^{-\beta^2 } }=\frac{\exp (-\lambda_0 
	\tau e^{-\beta^2 } )}{2\beta \tau }\left(1+\frac{1}{\lambda_0 
	\tau e^{-\beta^2 } }\right) \ ,
	\eqno(C14)
$$ 
$$
	\frac{d^2 }{d\tau^2 }\frac{\exp (-\lambda_0 \tau e^{-\beta^2 } 
	)}{2\lambda_0 \tau \beta e^{-\beta^2 } }=\frac{\lambda_0 
	e^{-\beta^2 } \exp (-\lambda_0 \tau e^{-\beta^2 } )}{2\beta 
	\tau }\left(1+\frac{2}{\lambda_0 \tau e^{-\beta^2 } }
	+O{\left(\frac{1}{\lambda_0 \tau }\right)}^2 \right) \ ,
	\eqno(C15)
$$
and then obtain,
$$
	N(\tau )\sim \lambda_0 \frac{e^{-\beta^2 } \exp (-\lambda_0 \tau 
	e^{-\beta^2 } )}{\sqrt{\pi }\beta \lambda_0 \tau 
	~{\mathrm{e}\mathrm{r}\mathrm{f}(\beta )}}\left(1+\frac{2}
	{\lambda_0 \tau e^{-\beta^2 } }\right) \ . 
	\eqno(C16)
$$
In the $\delta$-function limit we have $\sigma \to 0$ and $\beta \to \infty$, 
so that the second term in the expansion will dominate and $\alpha =2$,
consistent with the $\delta$-function limit we have seen as 
$p\to \infty$ in the polynomial case.  Comparing with the numerical solution,
it is apparent that a power-like form is found for 
$\lambda_0 \tau \approx 5-10$ 
and does not apply at larger values of $\tau$.  Note that
power laws are generally substantially different as $\sigma$ varies.  

\vskip1cm
{\sl Acknowledgements:}
We acknowledge valuable discussions with Dudok de Wit.
Part of the work was supported by NASA contracts 
NNG04EA00C and NNG09FA40C. JRJ and YIN acknowledge
NASA grant NNX16AQ87G.

\clearpage

\section*{ References }  

\def\ref#1{\par\noindent\hangindent1cm {#1}} 

\ref{Aschwanden, M.J. and McTiernan, J.M. 2010, ApJ 717, 683}
\ref{Aschwanden, M.J. 2011 {\sl Self-Organized Criticality in 
	Astrophysics.  The Statistics of Nonlinear Processes 
	in the Universe}, Springer-Praxis: New York, 416p.}
\ref{Aschwanden, M.J. 2012, A\&A 539:A2}
\ref{Aschwanden, M.J. and Freeland, S.M. 2012, ApJ 754:112}
\ref{Aschwanden, M.J. 2014, ApJ 782, 54}
\ref{Aschwanden, M.J. 2015, ApJ 814:19}
\ref{Aschwanden, M.J., Crosby, N., Dimitropoulouy, M., et al.~2016,
	SSRv 198:47}
\ref{Aschwanden, M.J. 2019a, ApJ 880, 105}
\ref{Aschwanden, M.J. 2019b, ApJ 887:57}
\ref{Aschwanden, M.J. and Dudok de Wit, T. 2021, ApJ 912:94}
\ref{Bak, P., Tang, C., and Wiesenfeld, K. 1987,
        Physical Review Lett. 59/27, 381}
\ref{Bak, P., Tang, C., and Wiesenfeld, K. 1988,
        Physical Rev. A 38/1, 364}
\ref{\bf Bauke, H. 2007, Eur.Phys.J.B. 58, 167}
\ref{Boffetta, G., Carbone, V., Giuliani, P., Veltri, P., 
	and Vulpiani, A. 1999, Phys.Rev.Lett. 83, 4662}
\ref{Bourouaine, S., Perez, J.C., Klein, K.C., Chen, C.H.K. 2020,
	ApJ 904, 308}
\ref{\bf Carbone, V., Cavazzana, R., Antoni, V., Sorriso-Valvo, L. et al.
	2002, Europhysics Letters 58/3, 349}
\ref{Cheng, Y., Zhang, G.Q., and Wang,F.Y. 2020, MNRAS 491, 1498}
\ref{Bronstein, I.N. and Semendjajev, A.K. 1960, Harry Deutsch:
	Z\"urich}
\ref{Dudok de Wit, P., Krasnoselskikh, V.V., Bale, S.D., et al. 
	2020, ApJSS  246:39}
\ref{Eastwood, J.P., Wheatland, M.S., Hudson, H.S., et al.
 	2010, ApJ 708, L95}
\ref{\bf Grandell, J. 1997, {\sl Mixed Poisson Processes},
	Chapman and Hall: London}  
\ref{Greco, A., Matthaeus, W.H., Servidio, S., Chuychai, P.
	and Dmitruk, P. 2009, ApJ 2, L111} 
\ref{\bf Goldsteinm, M.L., Morris, S.A., and Yen, G.G. 2004,
	Eur.Phys.J.B. 41, 255}
\ref{Gorobets, A. and Messerotti, M. 2012, SoPh 281, 651}
\ref{Grigolini, P., Leddon,D., and Scafetta,N. 2002,
 	Phys.Rev.Lett E, 65/4. id. 046203}
\ref{Guidorzi,C., Dichiara, S., Frontera,F., et al. 
 	2015, ApJ 801, 57}
\ref{Hawley, S.L., Davenport, J.R.A., Kowalski, A.F., Wisniewski, 
	John P. et al. 2014, ApJ 797, 12H} 
\ref{Hudson, H.S. 2020, MNRAS 491 4435}
\ref{\bf Kingman J.F.C. 1993, {\sl Poisson Processes},
	Oxford University Press}
\ref{Leddon, D. 2001, eprint arXiv:cond-mat/0108062, 
	Dissertation Abstracts International, Volume: 63-12, 
	Section: B, page: 5891; 56 p.}
\ref{Lepreti, F., Carbone, V., and Veltri,P. 2001, ApJ 555, L133}
\ref{Li,C., Zhong, S.J., Xu, Z.G., et al. 2018, MNRAS 479 L139}
\ref{Melatos, A., Peralta, C., and Wyithe, J.S.B. 2008, ApJ 672:1103}
\ref{Melatos, A., Howitt, G. 2018, ApJ 863, 196}
\ref{Morales, L.F. and Santos, N.A. 2020, SoPh 295, 155}
\ref{Newman, M.E.J. 2005, Contemp.Phys. 46, 323}
\ref{Norman, J.P., Charbonneau, P., McIntosh, S.W., and Liu, H.
 	2001, ApJ 557, 891}
\ref{Nurhan, Y.I., Johnson, J.R., Homan, R., et al.~(2021), GRL, (subm.)}
\ref{Sanchez,R., Newman, D.E., Ferenbaugh, W., et al.
 	2002, Phys.Rev E 66, 036124}
\ref{\bf Streit, R.L. 2010, {\sl Poisson Point Processes},
	Springer: New York}
\ref{Li, C., Zhong, S.J., Wang, L., et al. 2014, ApJ 792, L26}
\ref{Snelling, J.M., Johnson, J., Willard, J. et al. 
	2020, ApJ 869, 148}
\ref{Stumpf, M.P.H. and Porter, M.A. 2012, Science 335, 10 Feb 2012}
\ref{Uritsky, V.M., Paczuski, M., Davila, J.M., and Jones, S.I.
 	2007, Phys. Rev. Lett. 99, 025001}
\ref{Usoskin, I.G., Solanki, S.K., and Kovaltsov, G.A. 
	2007, A\&A 471, 301}
\ref{Wang, Y., Liu, L., Shen, C., et al. 2013, ApJ 763, L43}
\ref{Wang, F.Y., Dai, Z.G., Yi, S.X., and Xi, S.Q.
 	2015, ApJS 216, 8}
\ref{Wang, J.S., Wang, F.Y., and Dai, Z.G. 2017, MNRAS 471 2517}
\ref{Wanliss, J.A. and Eygand, J.M. 2007, GRL 34, 4107}
\ref{Wheatland, M.S., Sturrock,P.A., and McTiernan,J.M. 1998, 
	ApJ 509, 448}
\ref{Wheatland,M.S. 2000a, ApJ 536, L109}
\ref{Wheatland,M.S. 2000b, SoPh 191, 381}
\ref{Wheatland, M.S. 2001, SoPh 203, 87}
\ref{Wheatland, M.S. and Litvinenko, Y.E. 2002, SoPh 211, 255}
\ref{Wheatland, M.S. 2003, SoPh 214, 361}
\ref{Wheatland, M.S. 2004, ApJ 609, 1134}
\ref{Wheatland, M.S., and Craig, I.J.D. 2006, SoPh 238, 73}
\ref{Yeh, C.T., Ding, M., and Chen, P. 2005, ChJAA 5, 193}
\ref{Yi, S.X., Xi, S.Q., Yu, Hai et al. 2016, ApJS 224, 20}

\clearpage

\begin{table}[t]
\begin{center}
\normalsize
\caption{Power law slopes $\alpha$ 
of waiting time distributions, calculated by numerical simulations
for three model types (polynomial, sinusoidal, Gaussian),
as a function of the model exponent $p$. The theoretically predicted
slope is $\alpha=2+1/p$.}
\medskip
\begin{tabular}{ccccc}
\hline
        p   & $\alpha_P$ & $\alpha_S$ & $\alpha_G$ & $\alpha_{theo}$\\
            & Polynomial & Sinusoidal & Gaussian   & 2+1/p          \\  
\hline
\hline
   0.5 &   3.25 &   3.39 &   3.48 &   4.00\\
   1.0 &   2.99 &   3.02 &   2.98 &   3.00\\
   1.5 &   2.60 &   2.90 &   2.55 &   2.67\\
   2.0 &   2.51 &   2.61 &   2.34 &   2.50\\
   2.5 &   2.32 &   2.59 &   2.34 &   2.40\\
   3.0 &   2.30 &   2.56 &   2.33 &   2.33\\
   3.5 &   2.27 &   2.39 &   2.33 &   2.29\\
   4.0 &   2.25 &   2.39 &   2.33 &   2.25\\
   4.5 &   2.24 &   2.40 &   2.33 &   2.22\\
   5.0 &   2.15 &   2.40 &   2.24 &   2.20\\
   5.5 &   2.14 &   2.39 &   2.24 &   2.18\\
   6.0 &   2.13 &   2.37 &   2.25 &   2.17\\
\hline
\end{tabular}
\end{center}
\end{table}


\begin{table}
\begin{center}
\normalsize
\caption{Waiting time distributions measured from solar flares
over at least a half solar cycle ($T \gapprox 5$ years).}
\medskip
\begin{tabular}{llrllll}
\hline
Observations    &Observations     &Number         &Waiting        &WTD    &Powerlaw       &References\\
year of events & spacecraft or    &range          &time           &       &               &          \\
             & instrument      &               &$\tau    $     &       &$\alpha_{\tau}$ &     \\
\hline
\hline
1980-1989       & HXRBS/SMM       & 12,772        & $0.01-500$ hrs& PL    & $\approx 2.0$ & Aschwanden \& McTiernan (2010)\\
2002-2008       & RHESSI          & 11,594        & $2-1000$ hrs  & PL    & $\approx 2.0$ & Aschwanden \& McTiernan (2010)\\
1991-2000       & BATSE/CGRO      & 7212          & $1-5000$ hrs  & PL    & $2.14\pm0.01$ & Grigolini \etal (2002)\\
1975-1999       & GOES 1-8 A      & 32,563        & $1-1000$ hrs  & PL    & $2.4\pm0.1$   & Boffetta \etal (1999)\\
1975-1999       & GOES 1-8 A      & 32,563        & $1-1000$ hrs  & PL    & $2.16\pm0.05$ & Wheatland (2000a), Lepreti \etal (2001)\\
1975-2001       & GOES 1-8 A      & ...           & $1-1000$ hrs  & PL    & $2.2\pm0.1$   & Wheatland and Litvinenko (2002) \\
1996-2001       & SOHO/LASCO      & 4645          & $1-1000$ hrs  & PL    & $2.36\pm0.11$ & Wheatland (2003)\\
\hline
\end{tabular}
\end{center}
\end{table}

\clearpage


\begin{figure}		
\centerline{\includegraphics[width=1.0\textwidth]{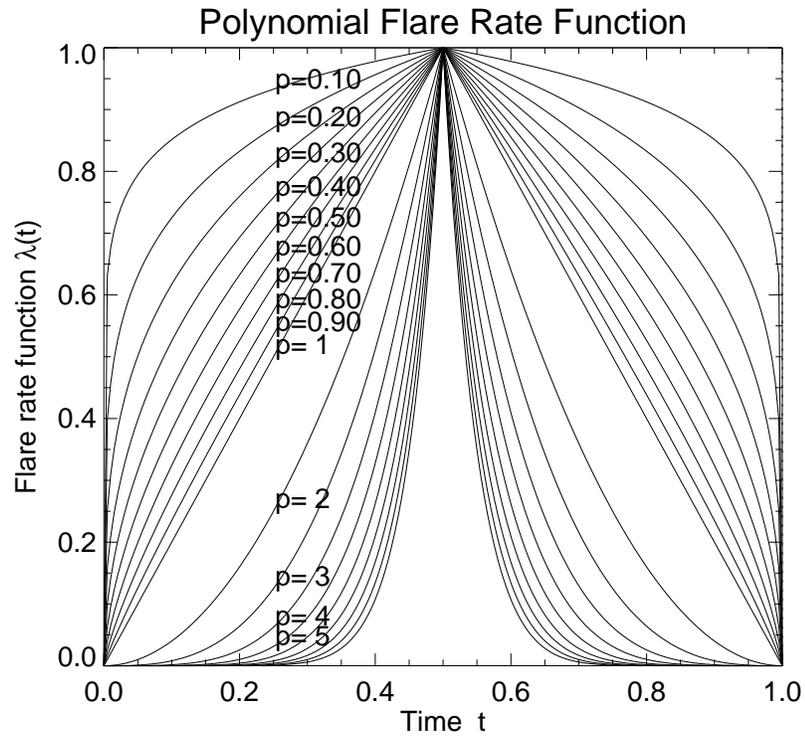}}
\caption{Parameter range of polynomial flare rate functions 
$\lambda(t)$ for exponents $p=0.1,...,10$, including the
linear case $p=1$ for $\lambda(t) \propto t$, or the
nonlinear (quadratic) case $p=2$ for $\lambda(t) \propto t^2$.}
\end{figure}

\begin{figure}		
\centerline{\includegraphics[width=1.0\textwidth]{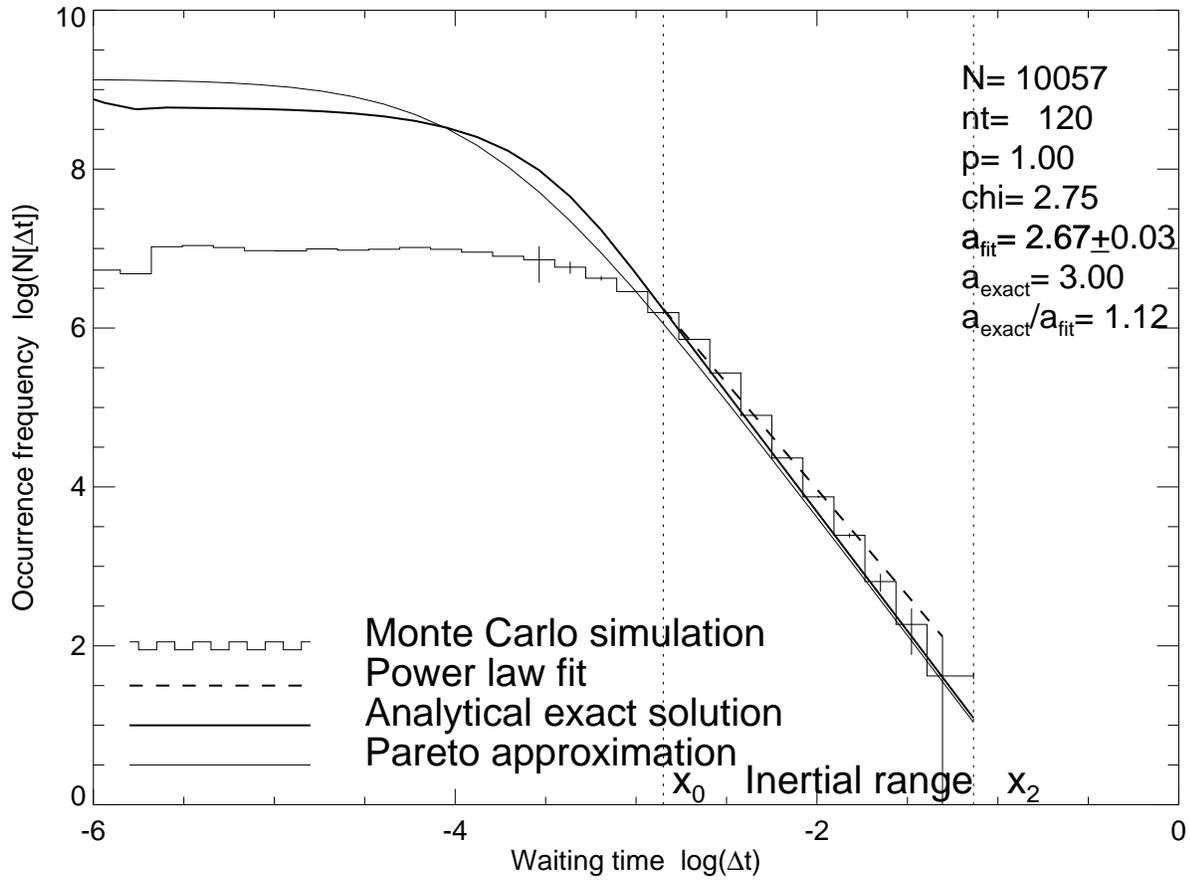}}
\caption{Comparison of three different calculation methods
of the waiting time distribution function: Monte Carlo 
simulation (histogram) with power law fit (dashed curve), 
analytical exact solution (thick solid curve),
and Pareto approximation (thin solid curve).
The fitting range is defined by the inertial range $[x_0, x_2]$.
The ratio of the slopes $\alpha_{exact}/\alpha_{fit}=1.12$ is
attributed to the Poissonian weighting bias.}
\end{figure}

\begin{figure}		
\centerline{\includegraphics[width=1.0\textwidth]{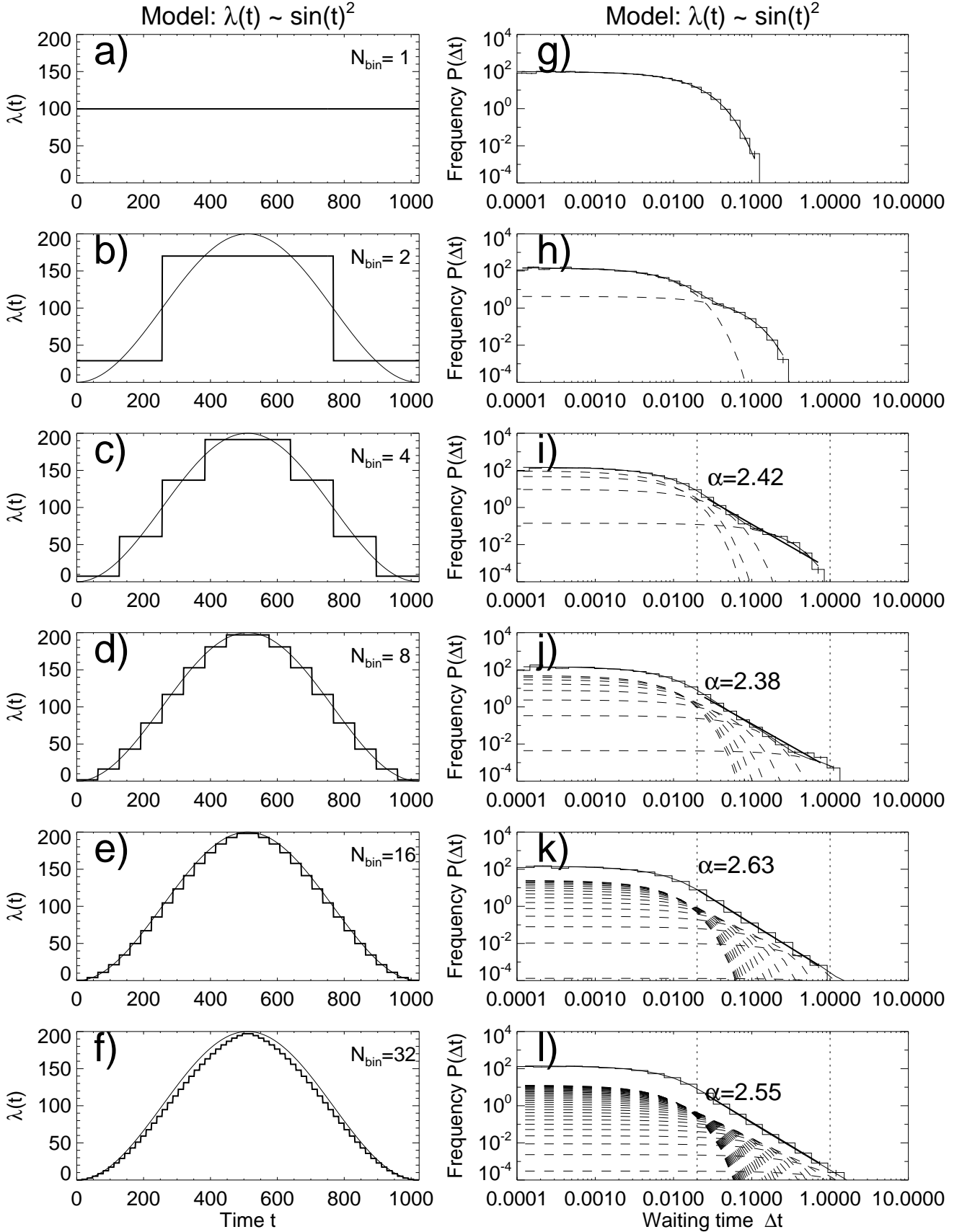}}
\caption{Six cases of numerical simulations of the 
waiting time distributions with different time resolutions
dt=(1 year)/$n_{bin}$, $n_{bin}=1,2,4,8,16,32$.
The waiting time components are shown individually for each bin 
(dashed curves in right panels), which sum up to the total
(histograms in right panels). Note that the power law slope
is almost constant and thus independent of the the time
resolution. The theoretical slope of a continuous sinusoidal
model with $\lambda(t) \propto \sin{(t)}^2$ with exponent
$p=2$ is $\alpha=2+1/p=2.5$ agrees well with the simulations
with $\alpha \approx 2.4-2.6$.}
\end{figure}

\begin{figure}		
\centerline{\includegraphics[width=1.0\textwidth]{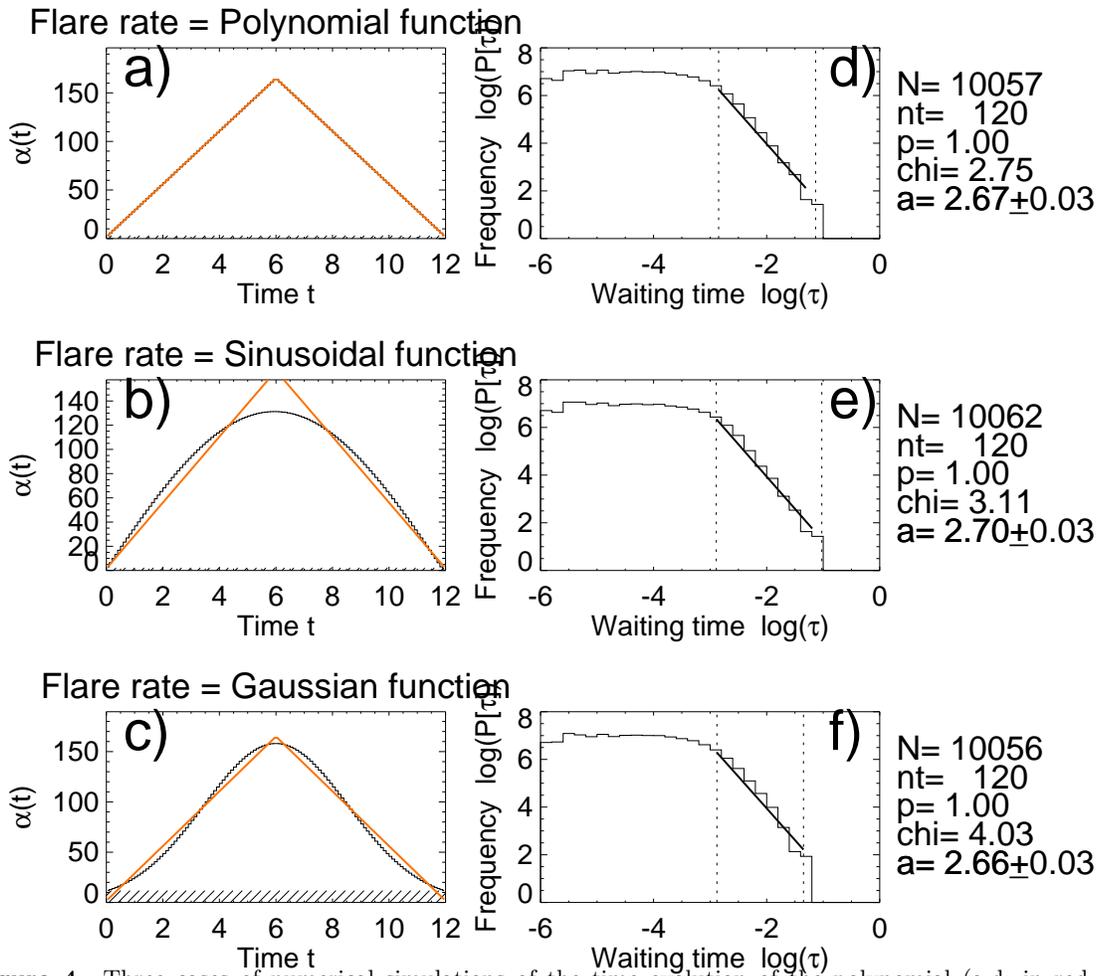}}
\caption{Three cases of numerical simulations of the time
evolution of the polynomial (a,d; in red color), 
the sinusoidal (b,e), and the Gaussian (c,f) flare rate 
function $f(t)$ (left panels a-c) are shown,
for linear flare rate changes ($p=1$).
The corresponding waiting time distributions 
(right panels d-f) are computed for the exponent $p=1$, 
and yield a best-fit power law slope of $a \approx 2.7$.
The power law slope values are not corrected for the Poissonian
weighting bias.} 
\end{figure}

\begin{figure}		
\centerline{\includegraphics[width=1.0\textwidth]{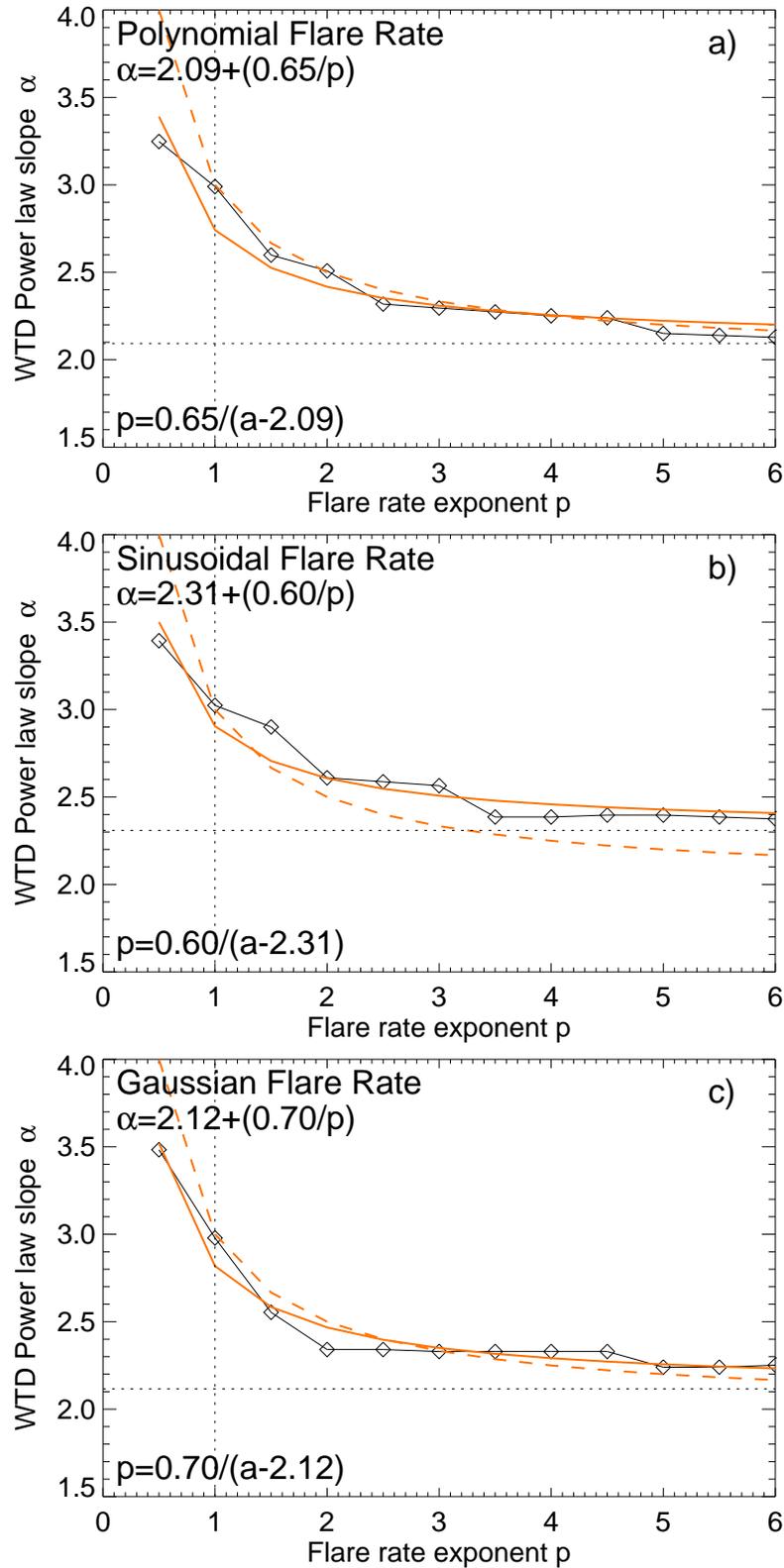}}
\caption{(a) Relationship between waiting time power law slope $\alpha$
and the flare rate function exponent $p$, computed by numerical
simulations (diamonds) for three different (polynomial, sinusoidal, 
Gaussian) flare rate model functions, and
fitted with the function $\alpha=\alpha_0+a_1/p$.
The theoretical prediction $\alpha=2+1/p$ is indicated
with dashed curves. The fitted power law slopes are corrected by
a factor of $q_{corr}=1.12$ (Fig.~2) to compensate for the
Poissonian weighting bias.}
\end{figure}

\begin{figure}		
\centerline{\includegraphics[width=1.0\textwidth]{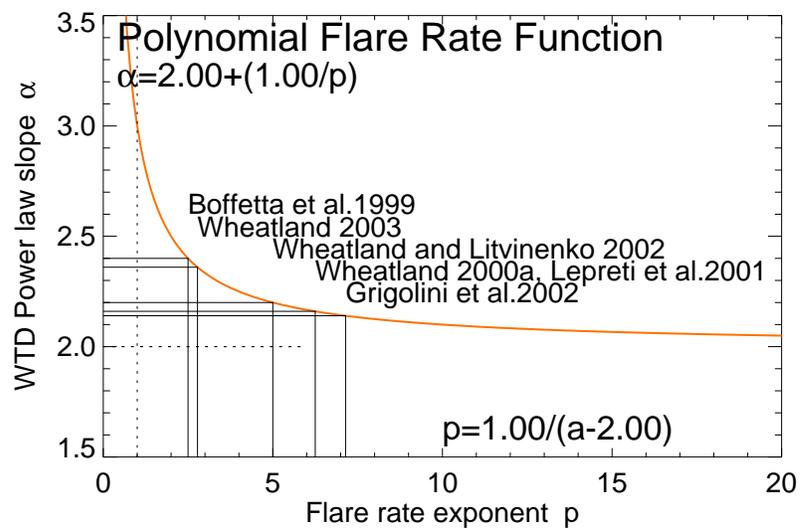}}
\caption{Inversion of the polynomial flare rate function 
exponents $p$ (x-axis) from the power law slope $\alpha$ 
(y-axis) of the theoretically predicted relationship
$\alpha=2+1/p$.}
\end{figure}

\begin{figure}		
\centerline{\includegraphics[width=1.0\textwidth]{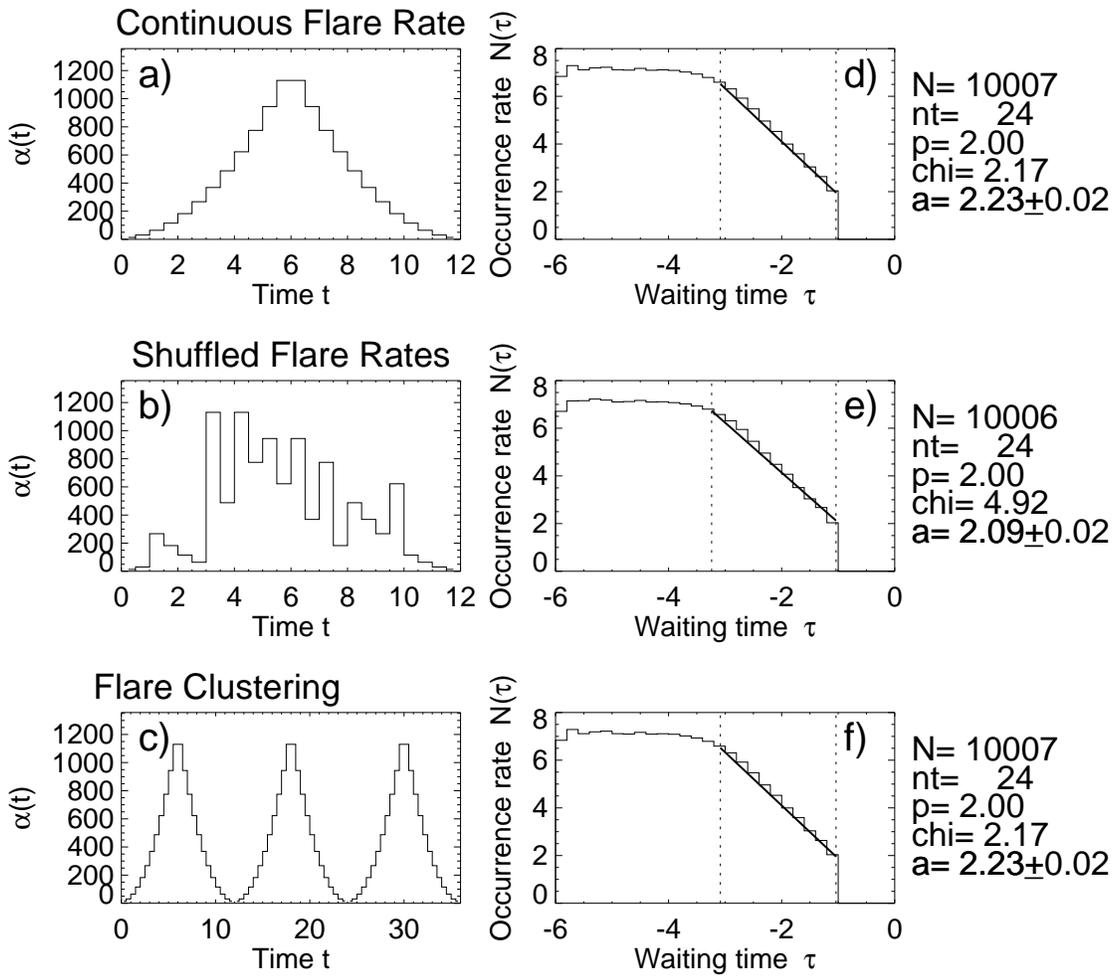}}
\caption{Comparison of three different flare rate functions
$\lambda(t)$: (a) polynomial (linear) function, 
(b) flucutations simulated by random shuffling of time bins,
and (c) flare clustering.
Note that the wait time distribution has near identical
values ($\alpha \approx 2.1-2.2$) due to the commutativity 
in summing the flare rates from different time bins.
The power law slope values are not corrected for the
Poissonian weighting bias.}
\end{figure}

\end{document}